\documentclass{emulateapj}
\usepackage{graphicx,amsmath,natbib}
\usepackage{epsf}
\usepackage{epstopdf}
\usepackage[normalem]{ulem}

\input{epsf}
\usepackage{epsfig}
\usepackage[countmax]{subfloat}
\usepackage{amsfonts}
\DeclareGraphicsExtensions{.jpg,.pdf,.png,.eps,.ps}
\usepackage{color}

\def\um{{\hbox {$\mu$m}}}
\def\lfir{{\hbox {$L_{\rm FIR}$}}}

\def\msol   {\ifmmode{{\rm M}_{\odot} }\else{M$_{\odot}$}\fi}
\def\lsol   {\ifmmode{{\rm L}_{\odot}}\else{${\rm L}_{\odot}$}\fi}
\def\sptOTFS {SPT0346-52}
\def\sptOFOE {SPT0418-47}
\def\sptOFTN {SPT0529-54}
\def\sptOFTE {SPT0538-50}
\newcommand{\hubble}{\textit{Hubble Space Telescope}}

\slugcomment{}
\shorttitle{ALMA Imaging of SPT Starburst Galaxies}
\shortauthors{Hezaveh et al.}
\begin{document}

\title{ALMA Observations of SPT-Discovered, Strongly Lensed, Dusty, Star-Forming Galaxies}

\def\McGill{1}
\def\Arizona{2}
\def\Davis{3}
\def\Caltech{4}
\def\UPenn{5}
\def\UChicago{6}
\def\ESO{7}
\def\CfA{8}
\def\Harvard{9}
\def\KICPChicago{10}
\def\EFIChicago{11}
\def\PhysicsUChicago{12}
\def\Miss{13}
\def\AAUChicago{14}
\def\ANL{15}
\def\Dal{16}
\def\Cambridge{17}
\def\NRAO{18}
\def\Berkeley{19}
\def\UFlorida{20}
\def\UCL{21}
\def\Colorado{22}
\def\LBNL{23}
\def\UCLA{24}
\def\ATNF{25}
\def\Michigan{26}
\def\MPIfR{27}
\def\Carnegie{28}
\def\STScI{29}
\def\CaseWestern{30}
\def\AstroMichigan{31}
\def\SAIC{32}
\def\Yale{33}
\def\IAS{34}

\author{
 Y.~D.~Hezaveh\altaffilmark{\McGill},
 D.~P.~Marrone\altaffilmark{\Arizona},   
 C.~D.~Fassnacht\altaffilmark{\Davis},
J.~S.~Spilker\altaffilmark{\Arizona},
J.~D.~Vieira\altaffilmark{\Caltech},
 J.~E.~Aguirre\altaffilmark{\UPenn},
 K.~A.~Aird\altaffilmark{\UChicago},
 M.~Aravena\altaffilmark{\ESO},
 M.~L.~N.~Ashby\altaffilmark{\CfA},
 M.~Bayliss\altaffilmark{\Harvard,\CfA}, 
B.~A.~Benson\altaffilmark{\KICPChicago,\EFIChicago}, 
L.~E.~Bleem\altaffilmark{\KICPChicago,\PhysicsUChicago}, 
 M.~Bothwell\altaffilmark{\Arizona},
M.~Brodwin\altaffilmark{\Miss},
J.~E.~Carlstrom\altaffilmark{\KICPChicago,\EFIChicago,\PhysicsUChicago,\AAUChicago,\ANL}, 
C.~L.~Chang\altaffilmark{\KICPChicago,\EFIChicago,\ANL}, 
S.~C.~Chapman\altaffilmark{\Dal,\Cambridge},
T.~M.~Crawford\altaffilmark{\KICPChicago,\AAUChicago}, 
A.~T.~Crites\altaffilmark{\KICPChicago,\AAUChicago},
C.~De~Breuck\altaffilmark{\ESO},
T.~de~Haan\altaffilmark{\McGill}, 
M.~A.~Dobbs\altaffilmark{\McGill}, 
E.~B.~Fomalont\altaffilmark{\NRAO},
E.~M.~George\altaffilmark{\Berkeley}, 
M.~D.~Gladders\altaffilmark{\KICPChicago,\AAUChicago}, 
A.~H.~Gonzalez\altaffilmark{\UFlorida}, 
T.~R.~Greve\altaffilmark{\UCL},	
N.~W.~Halverson\altaffilmark{\Colorado}, 
F.~W.~High\altaffilmark{\KICPChicago,\AAUChicago}, 
G.~P.~Holder\altaffilmark{\McGill}, 
W.~L.~Holzapfel\altaffilmark{\Berkeley}, 
S.~Hoover\altaffilmark{\KICPChicago,\EFIChicago}, 
J.~D.~Hrubes\altaffilmark{\UChicago}, 
K.~Husband\altaffilmark{\Cambridge},
T.~R.~Hunter\altaffilmark{\NRAO},
R.~Keisler\altaffilmark{\KICPChicago,\PhysicsUChicago}, 
A.~T.~Lee\altaffilmark{\Berkeley,\LBNL}, 
E.~M.~Leitch\altaffilmark{\KICPChicago,\AAUChicago}, 
M.~Lueker\altaffilmark{\Caltech}, 
D.~Luong-Van\altaffilmark{\UChicago}, 
 M.~Malkan\altaffilmark{\UCLA},
 V.~McIntyre\altaffilmark{\ATNF},
J.~J.~McMahon\altaffilmark{\KICPChicago,\EFIChicago,\Michigan}, 
J.~Mehl\altaffilmark{\KICPChicago,\AAUChicago}, 
 K.~M.~Menten\altaffilmark{\MPIfR},
S.~S.~Meyer\altaffilmark{\KICPChicago,\EFIChicago,\PhysicsUChicago,\AAUChicago}, 
L.~M.~Mocanu\altaffilmark{\KICPChicago,\AAUChicago},
 E.~J.~Murphy\altaffilmark{\Carnegie},
T.~Natoli,\altaffilmark{\KICPChicago,\PhysicsUChicago}, 
S.~Padin\altaffilmark{\Caltech,\KICPChicago,\AAUChicago}, 
T.~Plagge\altaffilmark{\KICPChicago,\AAUChicago}, 
C.~L.~Reichardt\altaffilmark{\Berkeley}, 
A.~Rest\altaffilmark{\STScI}, 
J.~Ruel\altaffilmark{\Harvard}, 
J.~E.~Ruhl\altaffilmark{\CaseWestern}, 
 K.~Sharon\altaffilmark{\KICPChicago,\AAUChicago,\AstroMichigan},
K.~K.~Schaffer\altaffilmark{\KICPChicago,\SAIC}, 
L.~Shaw\altaffilmark{\McGill,\Yale}, 
E.~Shirokoff\altaffilmark{\Caltech}, 
B.~Stalder\altaffilmark{\CfA}, 
Z.~Staniszewski\altaffilmark{\Caltech,\CaseWestern}, 
A.~A.~Stark\altaffilmark{\CfA}, 
K.~Story\altaffilmark{\KICPChicago,\PhysicsUChicago}, 
K.~Vanderlinde\altaffilmark{\McGill}, 
 A.~Wei\ss\altaffilmark{\MPIfR},
 N.~Welikala\altaffilmark{\IAS},
R.~Williamson\altaffilmark{\KICPChicago,\AAUChicago}
}

\altaffiltext{\McGill}{Department of Physics, McGill University, 3600 Rue University, Montreal, Quebec H3A 2T8, Canada}
\altaffiltext{\Arizona}{Steward Observatory, University of Arizona, 933 North Cherry Avenue, Tucson, AZ 85721, USA}
\altaffiltext{\Davis}{Department of Physics,  University of California, One Shields Avenue, Davis, CA 95616, USA}
\altaffiltext{\Caltech}{California Institute of Technology, 1200 E. California Blvd., Pasadena, CA 91125, USA}
\altaffiltext{\UPenn}{University of Pennsylvania, 209 South 33rd Street, Philadelphia, PA 19104, USA}
\altaffiltext{\UChicago}{University of Chicago, 5640 South Ellis Avenue, Chicago, IL 60637, USA}
\altaffiltext{\ESO}{European Southern Observatory, Karl-Schwarzschild Strasse, D-85748 Garching bei M\"unchen, Germany}
\altaffiltext{\CfA}{Harvard-Smithsonian Center for Astrophysics, 60 Garden Street, Cambridge, MA 02138, USA}
\altaffiltext{\Harvard}{Department of Physics, Harvard University, 17 Oxford Street, Cambridge, MA 02138, USA}
\altaffiltext{\KICPChicago}{Kavli Institute for Cosmological Physics, University of Chicago, 5640 South Ellis Avenue, Chicago, IL 60637, USA}
\altaffiltext{\EFIChicago}{Enrico Fermi Institute, University of Chicago, 5640 South Ellis Avenue, Chicago, IL 60637, USA}
\altaffiltext{\PhysicsUChicago}{Department of Physics, University of Chicago, 5640 South Ellis Avenue, Chicago, IL 60637, USA}
\altaffiltext{\Miss}{Department of Physics and Astronomy, University of Missouri, 5110 Rockhill Road, Kansas City, MO 64110, USA}
\altaffiltext{\AAUChicago}{Department of Astronomy and Astrophysics, University of Chicago, 5640 South Ellis Avenue, Chicago, IL 60637, USA}
\altaffiltext{\ANL}{Argonne National Laboratory, 9700 S. Cass Avenue, Argonne, IL, USA 60439, USA}
\altaffiltext{\Dal}{Department of Physics and Atmospheric Science, Dalhousie University, Halifax, NS B3H 3J5 Canada}
\altaffiltext{\Cambridge}{Institute of Astronomy, University of Cambridge, Madingley Road, Cambridge CB3 0HA, UK}
\altaffiltext{\NRAO}{National Radio Astronomy Observatory, 520 Edgemont Road, Charlottesville, VA 22903, USA}
\altaffiltext{\Berkeley}{Department of Physics, University of California, Berkeley, CA 94720, USA}
\altaffiltext{\UFlorida}{Department of Astronomy, University of Florida, Gainesville, FL 32611, USA}
\altaffiltext{\UCL}{Department of Physics and Astronomy, University College London, Gower Street, London WC1E 6BT, UK}
\altaffiltext{\Colorado}{Department of Astrophysical and Planetary Sciences and Department of Physics, University of Colorado, Boulder, CO 80309, USA}
\altaffiltext{\LBNL}{Physics Division, Lawrence Berkeley National Laboratory, Berkeley, CA 94720, USA}
\altaffiltext{\UCLA}{Department of Physics and Astronomy, University of California, Los Angeles, CA 90095-1547, USA}
\altaffiltext{\ATNF}{Australia Telescope National Facility, CSIRO, Epping, NSW 1710, Australia}
\altaffiltext{\Michigan}{Department of Physics, University of Michigan, 450 Church Street, Ann Arbor, MI, 48109, USA}
\altaffiltext{\MPIfR}{Max-Planck-Institut f\"{u}r Radioastronomie, Auf dem H\"{u}gel 69 D-53121 Bonn, Germany}
\altaffiltext{\Carnegie}{Observatories of the Carnegie Institution for Science, 813 Santa Barbara Street, Pasadena, CA 91101, USA}
\altaffiltext{\STScI}{Space Telescope Science Institute, 3700 San Martin Dr., Baltimore, MD 21218, USA}
\altaffiltext{\CaseWestern}{Physics Department, Center for Education and Research in Cosmology  and Astrophysics,  Case Western Reserve University, Cleveland, OH 44106, USA}
\altaffiltext{\AstroMichigan}{Department of Astronomy, University of Michigan, 500 Church Street, Ann Arbor, MI, 48109, USA}
\altaffiltext{\SAIC}{Liberal Arts Department, School of the Art Institute of Chicago,  112 S Michigan Ave, Chicago, IL 60603, USA}
\altaffiltext{\Yale}{Department of Physics, Yale University, P.O. Box 208210, New Haven, CT 06520-8120, USA}
\altaffiltext{\IAS}{Institut d'Astrophysique Spatiale, B\^atiment 121, Universit\'e Paris-Sud XI \& CNRS, 91405 Orsay Cedex, France}

\begin{abstract}
We present Atacama Large Millimeter/submillimeter Array (ALMA) 860 $\um$\ imaging of four high-redshift (z=2.8-5.7) dusty sources that were detected using the South 
Pole Telescope (SPT) at 1.4~mm and are not seen in existing radio to far-infrared catalogs. 
At 1.5\arcsec\ resolution, the ALMA
data reveal multiple images of each submillimeter source, separated by 1-3\arcsec, consistent with strong lensing by intervening galaxies
 visible in near-IR imaging of these sources. We describe a gravitational lens modeling procedure that operates on the measured visibilities and incorporates 
self-calibration-like antenna phase corrections as part of the model optimization, which we use to interpret the source structure.
 Lens models indicate that SPT0346-52, located at z=5.7, is one of the most luminous and intensely star-forming sources in the universe with a lensing corrected FIR luminosity of  $3.7 \times 10^{13} L_{\odot}$ and star formation surface density of 4200~\msol~yr$^{-1}$~kpc$^{-2}$.
We find magnification factors of 5 to 22, with lens Einstein radii of 1.1$-$2.0\arcsec\ and Einstein enclosed masses of 1.6-7.2$\times$10$^{11}$~\msol. 
These observations confirm the lensing origin of these objects,
 allow us to measure the their intrinsic sizes and luminosities, and demonstrate the important role that ALMA will play in the interpretation of lensed submillimeter sources.

\end{abstract}

\keywords{Gravitational lensing: strong ---
Galaxies: high-redshift ---
Techniques: interferometric
}

\section{Introduction}
Half of the energy produced by all objects in the history of the
universe has been absorbed and reemitted by dust \citep{dole06}. The Cosmic Infrared Background, first detected by the Cosmic Background Explorer satellite
\citep{puget96,hauser98,fixsen98}, is the aggregate emission from individual dusty galaxies across cosmic time \citep[e.g.,][]{lagache05}.
The brightest of these dusty star-forming galaxies (DSFGs) were discovered in deep submillimeter-wavelength images of the sky \citep{smail97,hughes98,barger98}, 
and have luminosities in excess of $10^{12}$~\lsol\ emitted primarily at rest wavelength in the far-infrared.
With star formation rates $>100-1000$~\msol~yr$^{-1}$, this population of DSFGs 
contributes a significant fraction of the total star formation density of the universe
at $z\sim2-3$, where their abundance peaks \citep[e.g.,][]{Chapman05}. These objects are the progenitors of the massive galaxies we observe today.

Despite the enormous total luminosity of the brightest DSFGs, their detection at submillimeter wavelengths requires lengthy exposures for ground-based 
facilities, and they are generally quite dim at optical/NIR wavelengths due to extinction. Studies of these objects and their extreme star formation
rates are limited by the observational costs of observing all but the brightest spectral lines and the poor spatial resolution achievable 
compared to the typical size of the star-forming regions. Gravitational lensing provides a solution to both of these problems, as has been 
demonstrated in a few spectacular cases \citep[e.g.,][]{kneib04,swinbank10,riechers11, fu12}. 
Lensed starburst galaxies can be examined at high spatial resolution
and with a more diverse set of diagnostics than the unmagnified population \citep{swinbank10}.

Predictions of a large population of gravitationally lensed, high redshift DSFGs \citep{blain96,negrello07} were recently verified by 
large-area millimeter/submillimeter surveys \citep{vieira10,negrello10,wardlow12}.
\citet{hezaveh11} also predicted the number counts of bright lensed objects for mm-wavelength surveys using a detailed numerical method, with a proper treatment of finite source effect and lens ellipticities, confirming that realistic lens models were able to match the observed number counts of dusty sources reported in \citet{vieira10}.
 These galaxies have a sky density of $\sim0.1$~deg$^{-2}$, 
and therefore can only be found in large numbers in extensive surveys. The South Pole Telescope (SPT; \citealt{carlstrom11}), which surveyed 2500 square degrees
to $\sim$~mJy depth at wavelengths of 3, 2, and 1.4 mm, has provided a sample of about one hundred candidate lensed sources \citep{vieira10}.
Initial investigations of these objects have found them to have properties consistent with unlensed starbursts, except for their large 
apparent luminosities \citep{greve12}. Morphological evidence of lensing cannot be discerned in data from the SPT survey or the 
single-aperture followup of \citet{greve12}, except in rare cases of lensing by clusters of galaxies, so arcsecond-resolution submillimeter imaging 
is required.

The Atacama Large Millimeter/submillimeter Array \citep[ALMA, ][]{Hills10}
 has begun operation in Chile, providing unprecedented submillimeter 
sensitivity even in early science. In this work, we employ ALMA to measure the arcsecond-scale structure of dusty extragalactic SPT sources at millimeter wavelengths, confirming the lensed nature of the four sources presented here. 
The observations reported here represent $<$10\% of our Cycle 0 sample and use only the compact configuration data, which was delivered first. Nevertheless,
from these ALMA data we are able to model the lensing geometry of these sources and de-magnify them, 
allowing them to be placed in the proper context within the high-redshift galaxy population. 
We are also able to infer the total mass and ellipticity of the lenses, a first step toward using the lensed submillimeter emission to 
characterize the lensing potential and its substructure. 
In Section 2 we describe the ALMA observations and supporting data 
and in Section 3 we describe a modeling technique 
for interferometric measurements of gravitationally lensed sources. Additional details on the integrated self-calibration step are 
included in the Appendix. In Section 4 we discuss the properties of the sources and lenses and present the conclusions in Section 5.
Throughout this work we assume a $\Lambda$CDM cosmology, with WMAP7 parameters, with $h=0.71$, $\Omega_M=0.27$, and $\Omega_\Lambda=0.73$ \citep{komatsu11}.

\section{Observations}
\subsection{ALMA Imaging}
The primary observations for this work were obtained from ALMA under a Cycle 0 program (2011.0.00958.S; PI: D. Marrone)
 in which 47 sources identified in the SPT survey are each observed in both the compact and extended array configurations.
The first data release for this program includes 20 sources observed in the compact array configuration, here we focus on 
the four sources for which these low-resolution data were sufficient to resolve the targets into multiple components. Of the 
remaining 16 sources, at least 8 are not point-like at the resolution of these observations, but we defer lens modeling for these sources 
until the remaining data are in hand.

The sources were targeted for brief snapshot observations with the dual-polarization Band 7 (275-373~GHz) receivers 
on two dates, 2011 Nov 16 and 28. 
The first local oscillator was set to 343.8~GHz, with all four spectral windows configured in time domain mode with 128 channels of 15.625~MHz
width centered at 5.125 and 7~GHz IF in each sideband. 
There were 16 and 14 antennas available on these days, respectively, arranged in a compact configuration.
The total elapsed time (for all sources, including those not published here) in the observations was approximately 4.1h.
The total integration time per source was 61 and 91 seconds in the first and second tracks, respectively, 
with 6.1-second sampling of the visibility data. The array alternately observed the science targets and gain calibrators (30~seconds), 
observing the calibrator every 3$-$4 minutes. Additional sources with known positions, precisely established against the 
International Celestial Reference Frame using very long baseline interferometry \citep{ma98}, were added to the tracks 
 to verify astrometry and calibration and observed with the same cycle as the science targets.

The flux scale was set with observations of Callisto on the first day.  On the second day the flux scale was derived by setting the flux of
quasar J0403-360 to 1.84~Jy, as reported by the ALMA staff from adjacent calibration observations. 
The absolute flux density scale is correct to within 15\%.
The antenna gains are equalized through gain calibration (amplitude and phase) on the main calibrators in each track. 
Short-timescale phase correction is achieved using the ALMA water vapor radiometry (WVR) system.
Very little variation is observed in these gain amplitudes through the tracks, and there is no evidence for 
atmospheric decorrelation on the longest baselines in the phase scatter of the target or calibrator visibilities after WVR phase correction.
The data were processed with the Common Astronomy Software Applications package \citep{mcmullin07,petry12} using standard steps 
for a continuum observation.

\begin{figure*}[th]
\begin{center}
\centering
\begin{minipage}[t]{0.45\linewidth}
\centering
\includegraphics[width=1.00\textwidth]{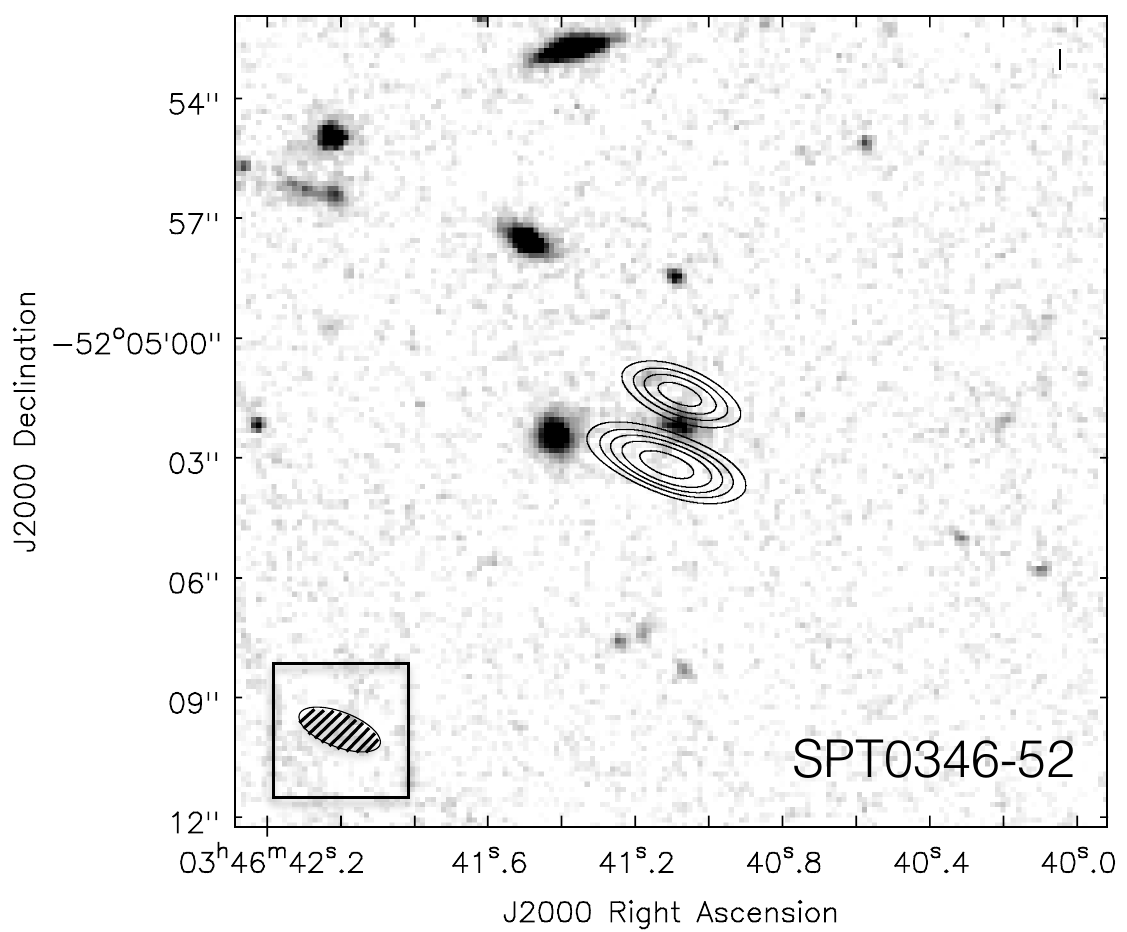}\\
\end{minipage}
\begin{minipage}[t]{0.45\linewidth}
\centering
\includegraphics[width=0.975\textwidth]{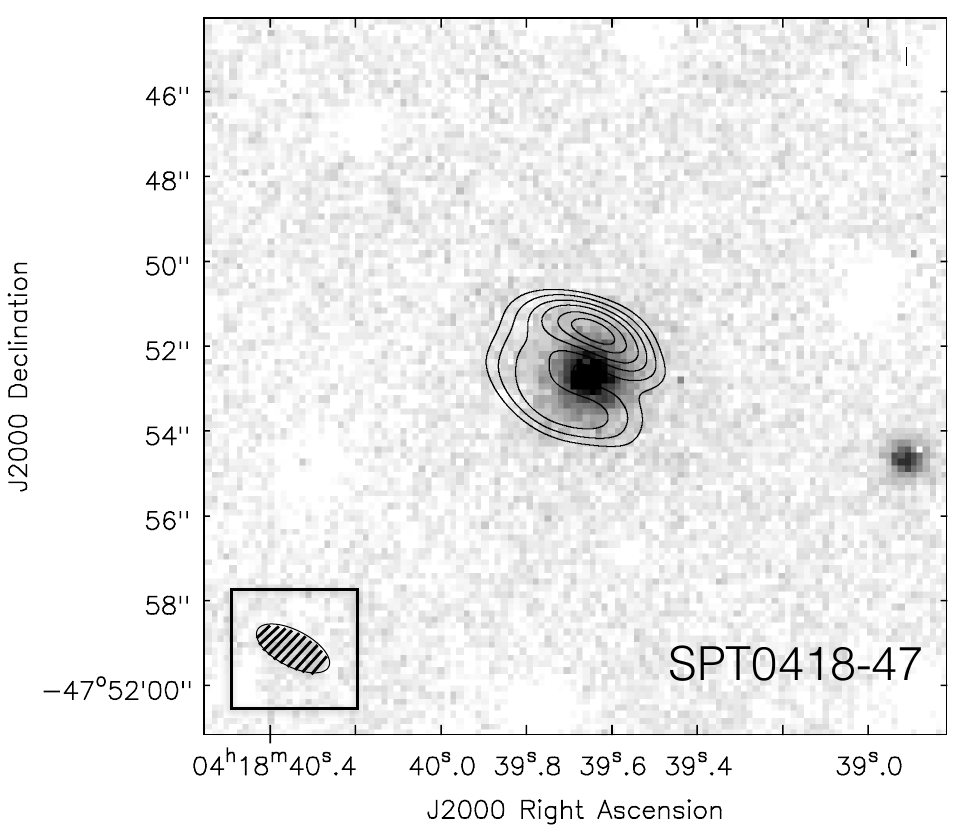}\\
\end{minipage}
\begin{minipage}[t]{0.45\linewidth}
\centering
\includegraphics[width=1.00\textwidth]{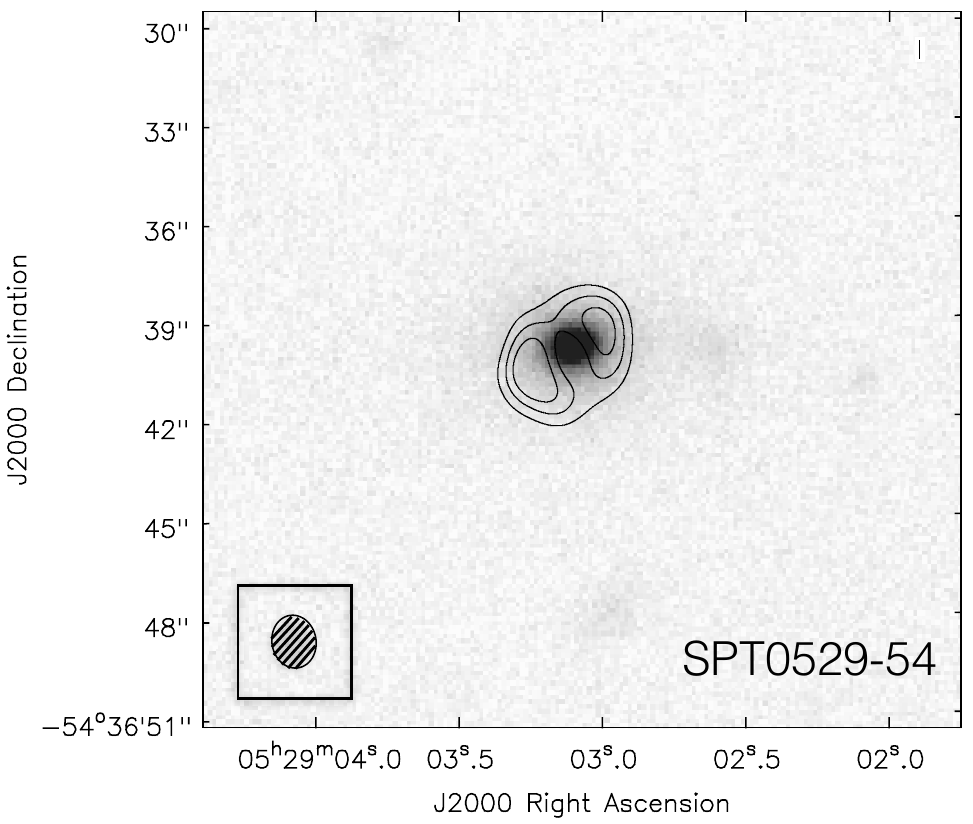}\\
\end{minipage}
\begin{minipage}[t]{0.45\linewidth}
\centering
\includegraphics[width=0.975\textwidth]{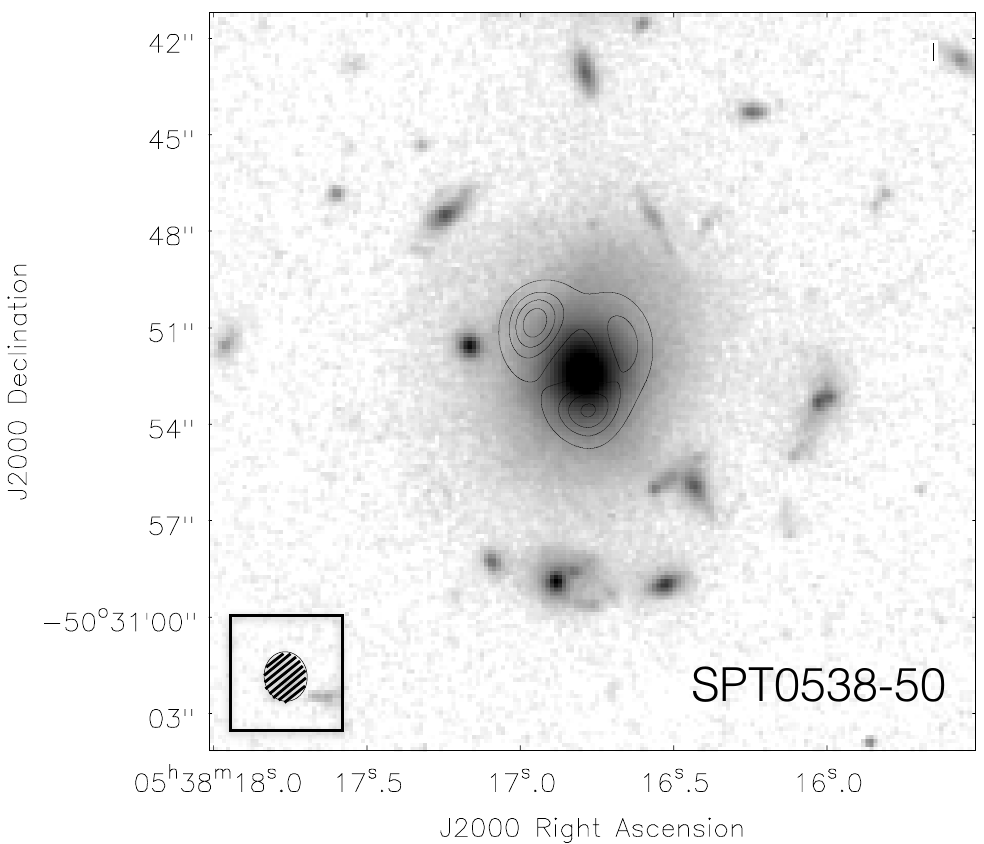}\\
\end{minipage}
\end{center}
\caption{\label{fig:images}
ALMA 350~GHz CLEANed images of the lensed DSFGs (contours), on top of NIR imaging of the galaxy field (greyscale) from
\hubble, \, VLT, and SOAR. Left to right they are \sptOTFS\ (\textit{HST}) and \sptOFOE\ (VLT) in the top row and \sptOFTN\ (SOAR) and \sptOFTE\ (\textit{HST}) in the bottom row. 
 The ALMA contours begin at 5$\sigma$, increasing in steps of 10$\sigma$ in the top left and 
bottom right images, and 5$\sigma$ in the other two, as in Figure~\ref{fig:modeling}.
The synthesized beam is shown in the lower left of each image.
The \textit{HST} images are a composite of the F110W and F160W filters, the VLT images are $K$ band from ISAAC, and the SOAR image is $K$ band from OSIRIS. }
\end{figure*}

The four well-resolved sources are listed in Table~\ref{table:sources}, we refer to them throughout the paper with shortened versions of 
their coordinate-based names (e.g., \sptOTFS\ for SPT-S J034640$-$5205.1).
Observations on Nov 16 included baselines of 15--150~k$\lambda$, resulting in synthesized beams of 
1.5\arcsec$\times$1.3\arcsec\ (FWHM) for the RA=5$^h$ sources. The $uv$ coverage on Nov 28 was less uniform and spanned 
15-240~k$\lambda$, providing a synthesized beam of 2.1\arcsec$\times$0.9\arcsec\ for the other two 
sources. Deconvolved source images and beam shapes are shown in Figure~\ref{fig:images}.

\begin{deluxetable*}{lcccccc|ccccc}
\tablecolumns{12}
\tablewidth{0pt}
\tablecaption{Source and Lens Parameters\label{table:sources}}
\tablehead{ & & & & & & & \multicolumn{5}{c}{Source Intrinsic Properties} \\
ID                  & $z_{S}$ & $z_{L}$ & $r_{\rm E}$ & $M_L$ 
& $\epsilon_L$ & $\mu$ & $R_{1/2}$ & \lfir & $\Sigma_{\rm FIR}$ & $S_{\rm 1.4 mm}$ & $S_{860 \mu\mathrm{m}}$  \\
(1) & (2) & (3) & (4) & (5) & (6) & (7) & (8) & (9) & (10) & (11) & (12)  }
\startdata
SPT-S J034640$-$5205.1 & 5.656 & \nodata &        1.124$\pm$0.004 & 3.73$\pm$0.04\tablenotemark{a} & 0.55$\pm$0.01 &  5.4$\pm$0.2 & 0.59$\pm$0.03 & 37.3$\pm$2.8 &   24 & 8.1 & 23.0 \\
SPT-S J041840$-$4752.0 & 4.224 & 0.265 &          1.390$\pm$0.012 & 2.39$\pm$0.04 & 0.20$\pm$0.03 & 21.0$\pm$3.5 & 1.07$\pm$0.17 &  3.8$\pm$0.7 & 0.74 & 1.6 & 5.0 \\
SPT-S J052903$-$5436.6 & 3.369 & 0.140 &          1.536$\pm$0.017 & 1.64$\pm$0.04 & 0.10$\pm$0.03 &  9.4$\pm$1.0 & 2.39$\pm$0.24 &  3.8$\pm$0.5 & 0.15 & 3.8 & 15.6 \\
SPT-S J053816$-$5030.8 & 2.782 & 0.404 &          1.987$\pm$0.009 & 7.15$\pm$0.05 & 0.13$\pm$0.02 & 20.5$\pm$4.0\tablenotemark{b} & \nodata &  4.5$\pm$0.9 & \nodata & 1.5 & 5.8 \\
\hspace{2em}\sptOFTE\ A& & & & & & 19.8$\pm$4.6 & 0.52$\pm$0.12 &  3.0\tablenotemark{c} &  2.4\tablenotemark{c} & & \\
\hspace{2em}\sptOFTE\ B& & & & & & 21.9$\pm$3.7 & 1.61$\pm$0.33 &  1.4\tablenotemark{c} & 0.12\tablenotemark{c} & & 
\enddata
\tablecomments{Column 1: SPT source name. Column 2: Background source redshift. Column 3: Lens redshift. Column 4: Einstein radius 
(arcsec). Column 5: Lens mass, interior to $r_E$ (10$^{11}$\msol). Column 6: Lens ellipticity. Column 7: Total magnification of 
background source. Column 8: Source radius, determined as the half-width at half maximum for the Gaussian source component in the 
model fit (kpc). Column 9: Intrinsic far infrared luminosity (10$^{12}$\lsol). Column 10: Intrinsic source flux (luminosity per area; 
$10^{12}$~\lsol/kpc$^{2}$). Column 11: Intrinsic 1.4~mm flux density, obtained as the SPT flux density divided by $\mu$ (mJy). 
Column 12: The same as (11), but for the ALMA 350~GHz (860~\um) flux density.
Parameter uncertainties do not include a contribution from the cosmological parameters. 
An additional 4\% uncertainty in mass is found by marginalizing over the WMAP7 parameter Markov chains.}
\tablenotetext{a}{Assuming that the lens is located at $z = 0.8$, see section~\ref{0346:lensredshift}.}
\tablenotetext{b}{Total magnification of the two source components, see section~\ref{sec:srcs}.}
\tablenotetext{c}{Derived assuming that \lfir\ is divided between components in the same ratio as the flux density in the model.}
\end{deluxetable*}

\subsection{Redshift Determinations}
Table~\ref{table:sources} shows source and lens redshifts, as available, for these four SPT sources. 
Redshifts for three of the dusty sources were obtained in another ALMA Cycle 0 project (2011.0.00957.S; PI: A. Weiss) through
a spectral line survey conducted using the Band 3 (84-116~GHz) receivers. 
 Complete results of this survey will be reported in \citet{vieira12} and \citet{weiss12}.
For each of these sources, multiple high-significance lines are detected in the Band 3 spectral scan, providing 
unambiguous redshifts.
In the case of \sptOFTE, a redshift was determined from a combination of millimeter-wavelength and optical spectroscopy, as 
described in \citet{greve12}, and the source was not included in the ALMA redshift search proposal.

The combination of NIR pre-imaging and submillimeter interferometric observations described below provided the basis for 
ground-based spectroscopic observations of the putative lens galaxies. 
The fields of \sptOTFS, \sptOFOE, and \sptOFTN\ were targeted first with $R$-band pre-imaging, then with multi-object masks, 
using the Mask Exchange Unit of FORS2 \citep{appenzeller98} on the Very Large Telescope (VLT) UT1, and exposing them for 
3$\times$900\,sec integrations each. The data were collected during February and March, 2012, in Service Mode 
(ESO program ID 088.A-0902) under an average seeing of $\sim1.1''$. The galaxies were observed with slits of $1''$ 
in width using the G300~V grism, yielding a velocity resolution of $\sim650$\,km~s$^{-1}$ or $\sim13$\AA, sampled 
at $\sim3.3$\AA\ pixel$^{-1}$.
\sptOFTE\ was observed with the XSHOOTER echelle spectrograph \citep{Vernet11} in longslit mode. 
For these observations, the slit widths were 1$''$ (UV-B), 0.9$''$ (VIS-R), and 0.9$''$ (NIR); the corresponding resolving 
powers are R = 5100, 8800, and 5600, sampled with 3.2, 3.0, and 4.0 wavelength bins respectively (after on-chip binning 
by a factor of two of the UV-B and VIS-R detectors). 6$\times$900\,sec integrations were obtained in February, 2010, however 
half of these were taken under worse conditions and we use only the better three integrations in the spectrum presented here.

\begin{figure*}[th]
\begin{center}
\centering
\begin{minipage}[t]{0.45\linewidth}
\centering
\includegraphics[width=1.0\textwidth]{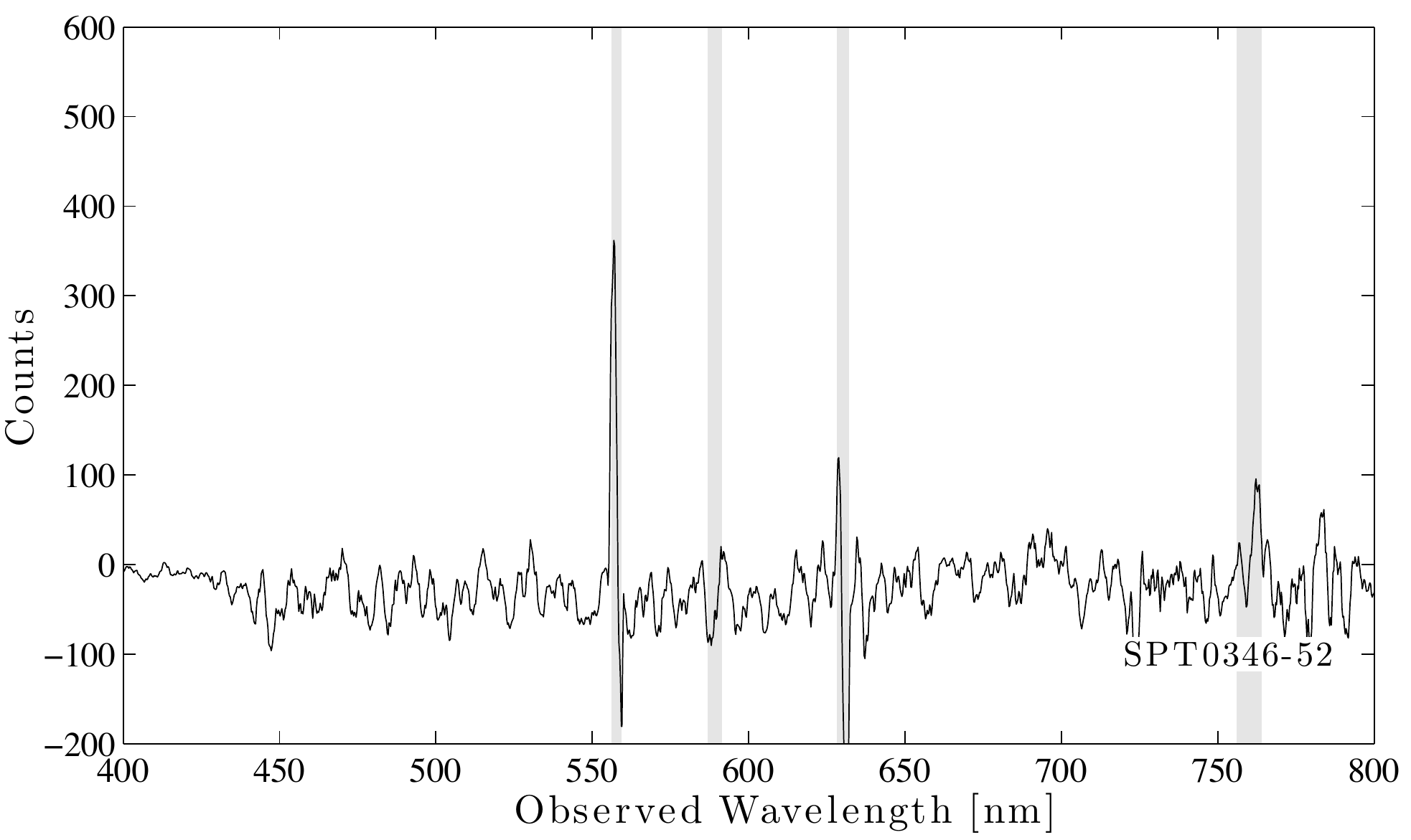}\\
\end{minipage}
\begin{minipage}[t]{0.45\linewidth}
\centering
\includegraphics[width=1.0\textwidth]{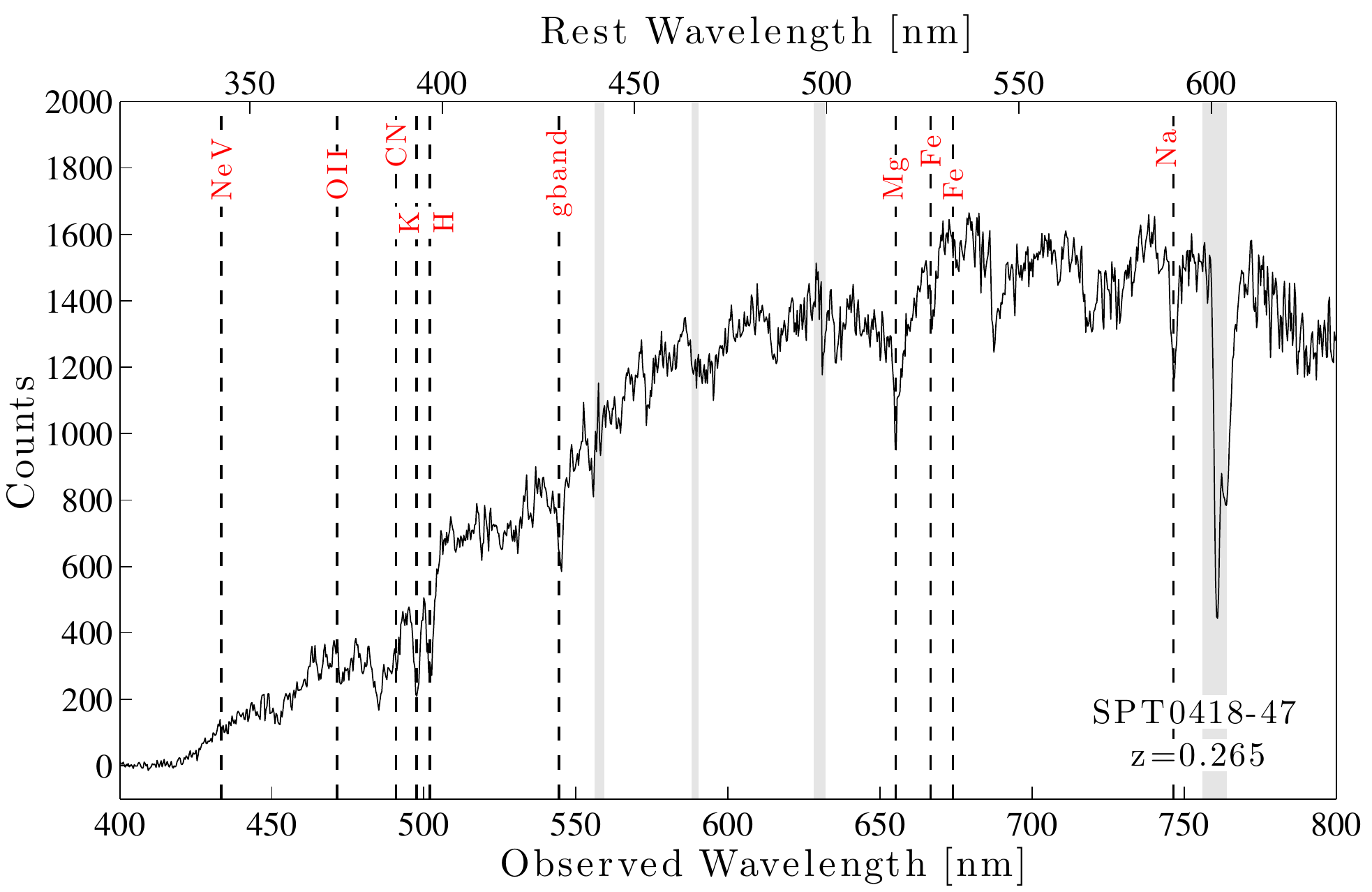}\\
\end{minipage}
\begin{minipage}[t]{0.45\linewidth}
\centering
\includegraphics[width=1.0\textwidth]{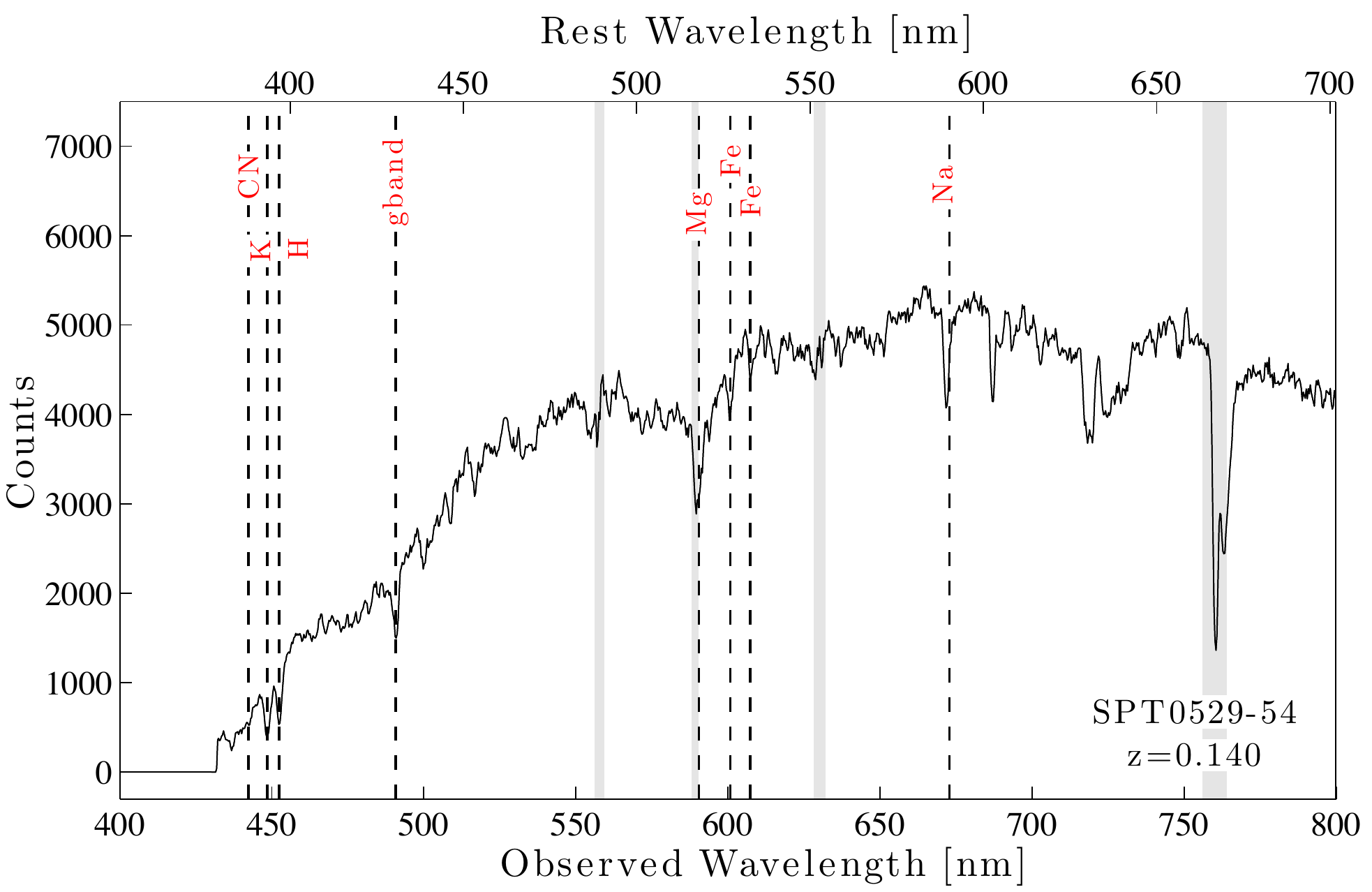}\\
\end{minipage}
\begin{minipage}[t]{0.45\linewidth}
\centering
\includegraphics[width=1.0\textwidth]{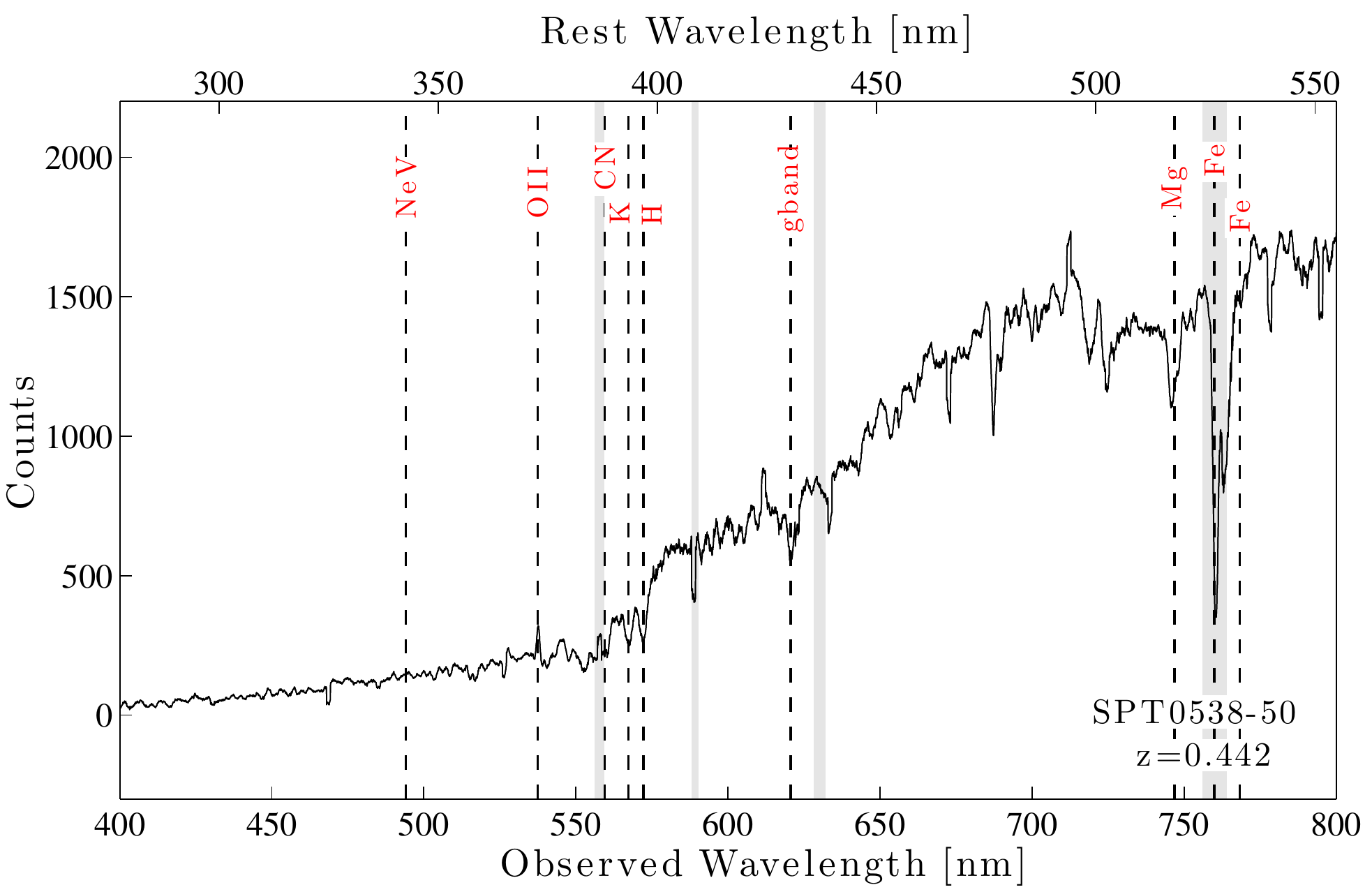}\\ 
\end{minipage}
\end{center}
\caption{\label{fig:spectra}
Optical spectra of the foreground lenses, from the VLT. The positions of major lines are marked by thin dashed lines. 
Sky lines are marked with grey shaded regions.} 
\end{figure*}

The observations were prepared and the data reduced using the standard ESO
pipeline\footnote{http://www.eso.org/sci/facilities/paranal/instruments/fors -- VLT-MAN-ESO-19500-1771}, 
performing bias and flat corrections, background subtraction, fringe correction, registration and combination, wavelength calibration,
and 1d spectral extractions. Spectra are shown in Figure~\ref{fig:spectra}. 
With the exception of \sptOTFS, redshifts for the lensing
galaxies were measured by fitting Gaussians to the 
Ca K+H absorption lines. 

\subsection{Other Supporting Data}
The small primary beam size of the 12m ALMA antennas at 350~GHz required that we improve upon the $\sim$10\arcsec\ positional
uncertainty of the SPT detections before proposing ALMA observations.
These targets were initially followed up with the Submillimeter Array (SMA; \citealt{Blundell04,Ho04}), or the Australia 
Telescope Compact Array\footnote{http://www.narrabri.atnf.csiro.au/} (ATCA; \sptOFTN\ only).
The SMA observations (Bothwell et al. in prep.) had typical angular resolution of 3\arcsec$\times$12\arcsec\ because of the low declination of the sources,
but those data gave some indication that the submillimeter emission was resolved. The 100~GHz ATCA detection of \sptOFTN\ was of 
low significance and did not reveal any ringlike structure. 

Deep near-infrared (NIR) imaging data were acquired from the Southern Astrophysical Research (SOAR) Telescope (OSIRIS; \citealt{depoy93}), the 
Very Large Telescope (VLT ISAAC; \citealt{moorwood98}), and the \hubble, as part of followup programs to examine lens properties and identify rest-frame optical emission 
from the background sources. 

The luminosities in Table~\ref{table:sources} are derived from photometric observations at millimeter to submillimeter wavelengths.
In addition to SPT photometry (1.4 and 2~mm), we use 870~\um\ observations from LABOCA \citep{siringo09} on the APEX telescope (in Max Planck time) and 
{\it Herschel}-SPIRE photometry at 250, 350, and 500~\um. The LABOCA data were analyzed according to the procedure 
described in \citet{greve12}. The {\it Herschel} data are reduced as described in \citet{weiss12}. All photometric data, including 
SPT measurements and the data previously reported in \citet{greve12} for \sptOFTN\ and \sptOFTE, are provided in \citet{weiss12}.
The LABOCA measurements of the total 870~\um\ flux density agree with the 
total flux density in the modeled ALMA data (described below) to within calibration uncertainties, typically 10\%.
Luminosities are calculated using a greybody model like that described in \citet{greve12}, though in the present case the $\tau=1$ 
wavelength is a parameter of the fit. The effect of freeing this parameter is to broaden the peak of the spectral energy distribution (SED) 
as needed to match the short-wavelength {\it Herschel} data points.

\section{Results and Analysis}
Figure~\ref{fig:images} presents the resolved structure of the 350~GHz emission associated with the SPT sources,
as revealed in Cycle 0 ALMA observations, and the NIR emission at the same positions.
In all cases there is a clear NIR counterpart to the submillimeter source, though with no structural correspondence
between the infrared sources and the emission found in the ALMA images. The morphologies of these sources at submillimeter wavelengths are indicative of gravitational lensing. 
The redshift identifications reported in Table~\ref{table:sources}, with very high redshifts measured for the dusty emission behind 
low-redshift foreground galaxies, clearly confirm these systems as galaxy-galaxy lenses. A similar finding was reported in 
\citet{negrello10}, where bright sources selected at much shorter wavelength from wide-field submillimeter surveys 
with {\it Herschel}-SPIRE were also confirmed to be gravitationally lensed.

\subsection{Lens Modeling}
The ALMA interferometer measures visibilities, Fourier components of the sky intensity distribution across a two-dimensional range of 
spatial frequencies, rather than directly imaging the emission. 
To properly compare these data with a source model, we must perform our analysis in the visibility plane, where the measurement
and its noise are well understood. Past techniques for modeling interferometric lens observations \citep[e.g.,][]{wucknitz04} have generally 
operated on reconstructed images that are subject to difficult-to-model biases and noise properties. Furthermore, residual errors
in calibration in the visibility data, such as those arising from imperfect knowledge of the antenna positions or uncompensated atmospheric 
delay, are often corrected as part of the imaging process through an iterative clean/self-calibration technique \citep{cornwell99}. However, the inclusion of the cleaning step in the determination of these
corrections, which are then applied to the visibility data themselves, changes the data in ways that are not easily included in the modeling uncertainties. Here we describe a 
visibility-based lens modeling technique that simultaneously determines these self-calibration phases so that we incorporate the full
range of uncertainty present in the measurements.

We model the lenses by generating model lensed images that are subjected to simulated observations and compared to the data. 
The source is assumed to have a symmetric Gaussian light profile with four free parameters: flux density F$_S$,  radius R$_S$, 
and positional offsets from the lens center $X_S$, $Y_S$.  The lens is modelled as a Singular Isothermal Ellipsoid (SIE) with five 
free parameters: mass inside the Einstein radius $M_L$, ellipticity $\epsilon_L$, orientation angle $\theta_L$  (east of north), and position $X_L$, $Y_L$. The SIE profile has been shown to be a good approximation to galaxy-scale density profiles \citep{treu04,koopmans06,koopmans09}. 
Since lensed image positions provide very precise measurements of the projected mass interior to the images we report this robustly measured quantity in Table~\ref{table:sources}, the total mass ($M_L$) inside the Einstein radius ($r_E$) of the lensing galaxies. However, the total halo masses associated with these galaxies will be much higher. Magnification, $\mu$, is calculated as the ratio of the total lensed to unlensed flux.

Given a set of lens and source parameters we make a high-resolution image of the lensed source, which we pad with zeroes and
Fourier transform to generate model visibilities. For each ALMA visibility, we interpolate the Fourier transformed image to the ALMA $u$, $v$ 
coordinates. We also correct for the antenna primary beam attenuation by multiplying the sky model images by the primary beam pattern before sampling the Fourier modes. We use a symmetric Gaussian with FWHM of 18\arcsec\ for the primary beam, which leads to minimal attenuation for these 
well-centered sources.

\begin{figure*}[ht]
\centering
\makebox[0cm]{\includegraphics[width=16cm]{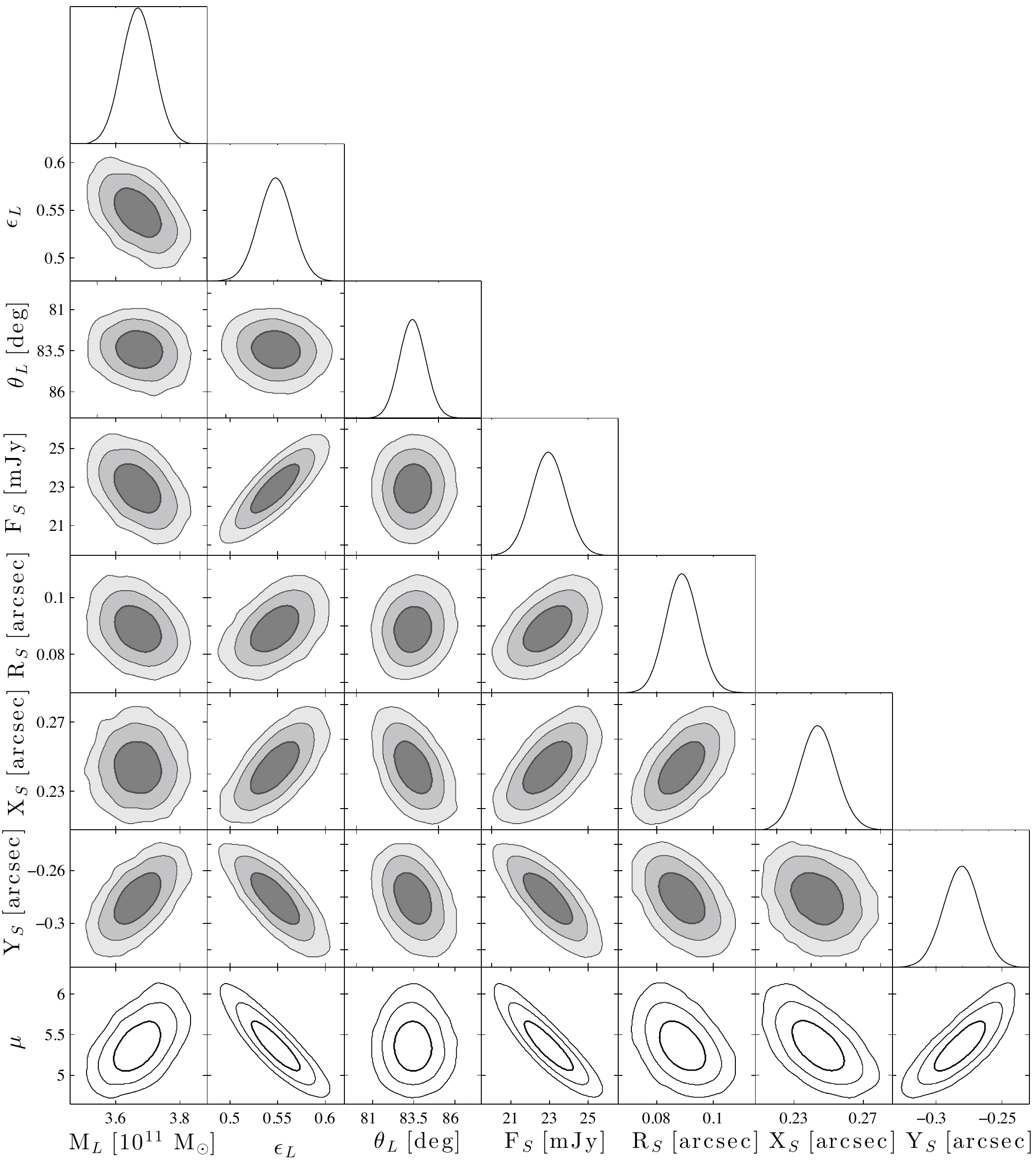}}
\caption{Parameter degeneracy plot for \sptOTFS, a representative example of the model uncertainties. 
Some nuisance parameters are not shown. 
Marginalized distributions for each parameter along the bottom axis are shown as histograms on 
the diagonal. The bottom row shows correlations between $\mu$, which is derived from the model and not a fit parameter, and the 
model parameters.  The contours show the 1, 2, and 3 $\sigma$ confidence regions.
\label{fig:triangle}}
\end{figure*}

The agreement between data and model visibilities is determined by calculating the $\chi^2$ between them, which requires an estimate of the visibility noise.
Because we observe strong sources, the visibility scatter has a contribution from the sky signal and
cannot be used to determine the noise level. 
Instead, we derive noise levels by measuring visibility scatter after differencing 
visibilities that are adjacent in time for the same baseline/polarization/IF, which has the effect of removing all sky signal. 
This gives results that are identical to those found by scaling the visibility noise to obtain a reduced $\chi^2$ of unity for the best-fit models.
We explore the model parameter space using a Markov Chain Monte Carlo (MCMC) method with  
Metropolis-Hastings sampling. An example of the parameter degeneracies in these fits is shown in Figure~\ref{fig:triangle}.

An additional complication of the modeling procedure is the presence of uncorrected antenna-based phase errors in the data. 
Observations of test sources, quasars with positions referenced to the International Celestial Reference Frame (ICRF), show significant residual phase errors after the 
primary phase calibration step. For simple source structures, self-calibration using CLEAN components as input for the phase
correction is a standard procedure to improve image fidelity \citep[e.g.,][]{taylor99}. However, in this application it would add significant 
complication to model the clean/self-cal process as part of the fitting, and leaving the phase errors uncorrected can significantly bias
the model parameters because of the strong sensitivity to image flux ratios in the model. We have therefore developed a procedure
to determine the self-cal phases as part of the model fitting. We optimize the $\chi^2$ for each step in the Markov chain by adjusting
the $N-1$ antenna phases. We find that the resulting phases vary little over the chain and closely resemble those found for nearby point sources 
added to the tracks to test the calibration and astrometry, giving us confidence in this method. Additional details of the method and 
simulations of its effectiveness are provided in the Appendix.

The model parameter space is complex, with the possibility of multiple isolated minima separated by high barriers in $\chi^2$. 
To decrease the possibility of missing important minima in the posterior we search the space more broadly by ``tempering'', which is similar to the simulated annealing method \citep[e.g.,][]{press07}. A control parameter, T (the analog of temperature), is introduced to flatten the posterior surface and to make the minima more accessible. This is achieved by raising the posterior to the power 1/T (with T$>$1).

\section{Discussion}
\label{sec:srcs}
A primary goal of this work is to determine the lensing configuration for bright SPT starburst galaxies and derive the total magnification. Using a simple multi-component 
source model and a range of lensing geometries, \citet{hezaveh12a} showed that differential magnification of a DSFG 
can distort the SED in unpredictable ways by differentially lensing different source plane regions (see also \citealt{serjeant12} for a similar effect concerning molecular line ratios).
A lens model is therefore essential if we wish to correctly interpret observations of these targets and place them in context with existing samples of unlensed starbursts.
 In principle, spatially resolved imaging of each molecular line is required to map that particular line to the source plane, however a single lens model based on continuum imaging, combined with physically motivated models for the relative filling factors of the emitting regions in other bands or molecular lines, can be used to place limits on the differential magnification between components.

\begin{figure*}[ht]
\begin{center}
\centering
\begin{minipage}[t]{1.0\linewidth}
\centering
\includegraphics[width=1.00\textwidth]{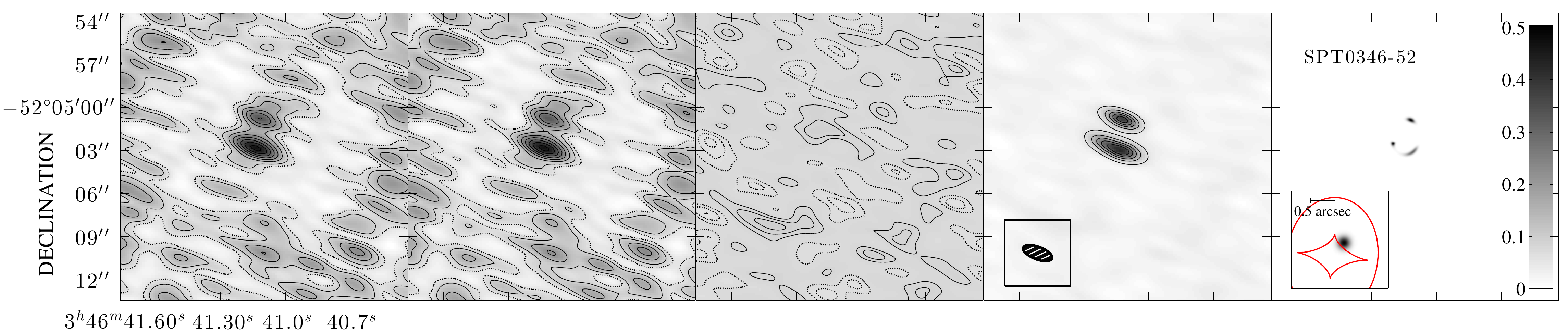}\\
\end{minipage}
\begin{minipage}[t]{1.0\linewidth}
\centering
\includegraphics[width=1.00\textwidth]{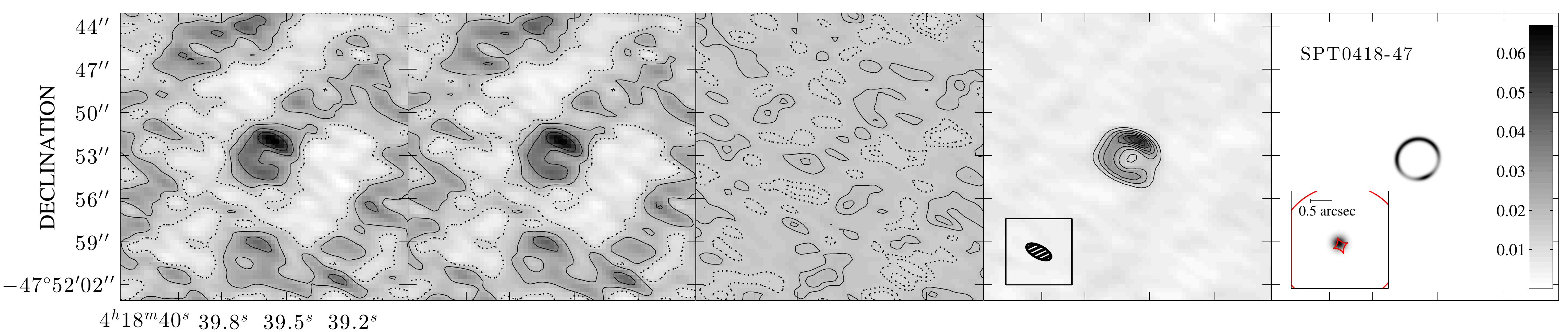}\\
\end{minipage}
\begin{minipage}[t]{1.0\linewidth}
\centering
\includegraphics[width=1.00\textwidth]{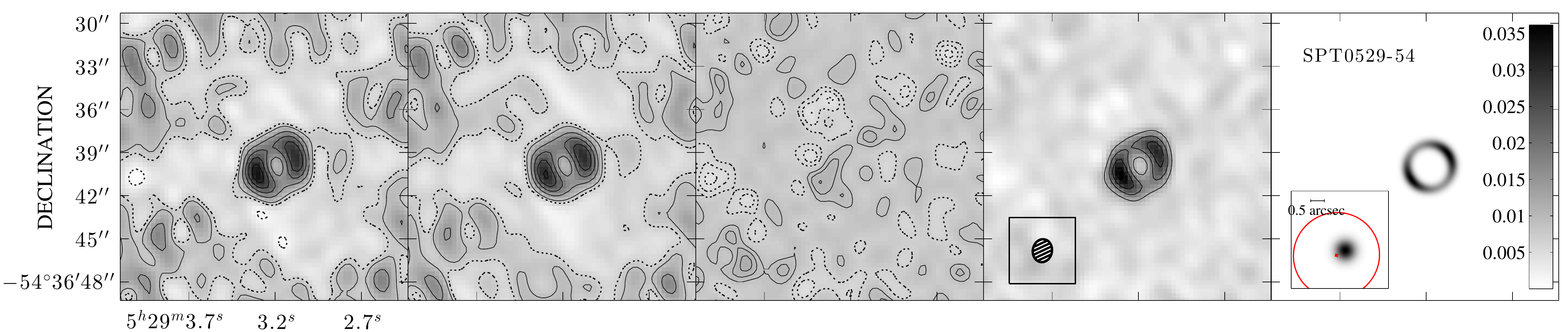}\\
\end{minipage}
\begin{minipage}[t]{1.0\linewidth}
\centering
\includegraphics[width=1.00\textwidth]{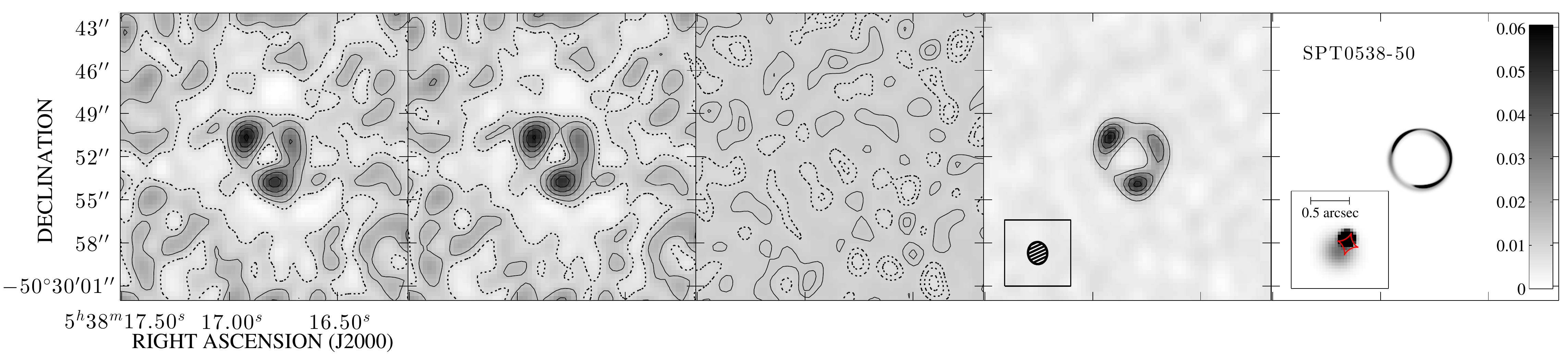}\\
\end{minipage}
\end{center}
\caption{\label{fig:modeling}
Modeling of the observed submillimeter emission. In each row, the panels from left to right are: the ``dirty'' image, 
model dirty image, residuals after subtraction, CLEANed image, and fully resolved lens model. 
Contours in the left two columns start at $\pm$3 times the
rms noise in the residual map (third column) and increase in steps of 10 times the rms noise, except in the second row where they increase in 
steps of 5 times the rms noise. 
Contours in the residuals are $\pm$ 1, 2, etc., times the rms noise, which is (top to bottom) 1.5, 1.7, 1.0, and 1.0~mJy.
The contours in the CLEANed images start at 5 $\sigma$ and increase in levels of $10 \sigma$ for the top and bottom rows and $5 \sigma$ in the two middle rows.
The apparently high significance structure in the dirty maps away from the source is due to the sidelobes of the 
synthesized (or ``dirty'') beam, and should therefore be reproduced by the model in the second panel.
The insets in the last panels show a magnified view of the positions of the 
emission in the source plane (greyscale) relative to the lensing caustics (red).
In the source model for \sptOFTE\ the greyscale is 
truncated at 10\% of the peak intensity to make the second source component visible. The intensity scale for the right panel, 
in Jy~arcsec$^{-2}$, is given by the colorbar in each figure.}
\end{figure*}

\begin{figure}[h]
\plotone{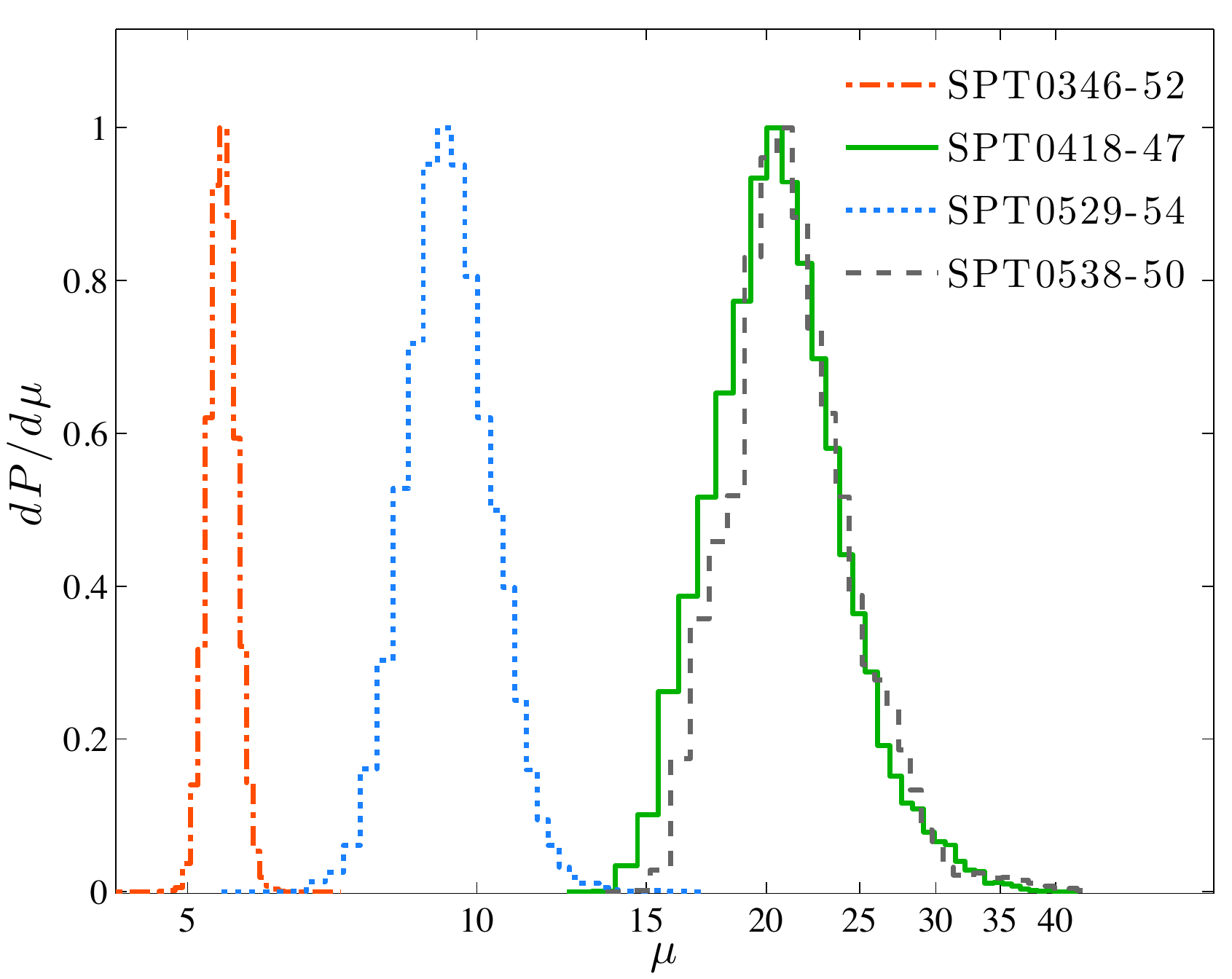}
\caption{\label{fig:mu_dist} 
The source magnification distributions derived from the lens models. 
 In the case of \sptOFTE, the total magnification 
of the two source-plane components is shown.}
\end{figure}

Models for the four sources are shown in the rightmost panels of Figure~\ref{fig:modeling}. Key parameters of the models are reported
in Table~\ref{table:sources}. The lens models permit the calculation of some intrinsic properties of the lensed galaxies by 
simply dividing observed properties by the magnification. The four 
sources presented here were detected with 1.4~mm flux densities ranging from 30$-$45~mJy.
By correcting for the lensing magnification (Figure~\ref{fig:mu_dist}) we find that the intrinsic 1.4 mm and 860~\um\  fluxes of these sources vary by a factor of several.
The 860~\um\ fluxes span the range 
for sources identified in blank field surveys with SCUBA \citep[e.g.,][]{greve04,coppin06,scott08,weiss09}, suggesting that even without the lensing boost the 
galaxies presented in this work would still be identified as luminous starbursts. The far-IR luminosities are even more widely varying, though
always ultra luminous infrared galaxy (ULIRG) class, indicating star formation rates of several hundred to several thousand \msol~yr$^{-1}$. 

The intrinsic sizes of the source-plane emission regions ($R_{1/2}$ in Table~\ref{table:sources}, the HWHM for the Gaussian model components) range from 0.5 to 2.4~kpc. 
This is at the lower range of the sizes generally inferred for starbursts from a variety of observables 
\citep[e.g.,][and references therein]{bothwell10, rujopakarn11, daddi10, tacconi06, tacconi10}, 
including synchrotron emission \citep{chapman04,biggs08} low-J \citep{ivison11} and even higher-J \citep[e.g.,][]{tacconi06} CO measurements. 
However, the selection of strongly lensed galaxies can result in a biased distribution of intrinsic source size, as 
noted by \citet{hezaveh12a}.  

Some comments on the individual sources:

{\it \sptOTFS}: This source has the lowest magnification of the set (5.4$\pm$0.2), and being the brightest of the four targets in apparent $S_{\rm 1.4mm}$, 
it is by far the most luminous after correction for magnification. Unlensed, with a flux density of 26~mJy it would be among the 860~\um-brightest high-redshift galaxies known. 
Given the source size, the flux (luminosity per area) is 
$2.4\times10^{13}$~\lsol~kpc$^{-2}$, implying a star-formation rate of 4200~\msol~yr$^{-1}$~kpc$^{-2}$ assuming a standard conversion 
for starbursting galaxies \citep{kennicutt98}. This is 
remarkably high, 50$\times$ higher than the average value found in marginally resolved starbursts by \citet{tacconi06} and a factor of several
higher than observed for individual GMCs in the highly magnified source SMM~J2135-0102 \citep{swinbank11}.
The flux is comparable to the Eddington limit \citep{thompson05}, and the degeneracy between $\mu$ and R$_S$ 
(shown in Figure~\ref{fig:triangle}) is such that larger magnification decreases R$_S$ and further increases the flux. 

\citet{thompson05} noted that such fluxes are relatively 
common among low-redshift ULIRGs, which are generally
more compact than high-redshift galaxies of similar luminosity,
suggestive of a self-regulating process. Because we find the luminosity of \sptOTFS\ to arise in a very compact 
region, its mode of star formation may be more similar to lower-redshift ULIRGs than to most other high-redshift starbursts, despite being at 
$z=5.7$. \citet{walter09} found a similar star formation density over a region of similar size in a $z$=6.4 quasar, so this is not unprecedented 
in the early universe.
The optical spectrum of the lens does not show any lines or features that result in a robust redshift for the lens. However, the only lens modeling parameter that is degenerate with the redshifts is the lens mass. The mass of this lens is reported assuming $z=0.8$ for the lens.   \label{0346:lensredshift}

{\it \sptOFOE} and {\it \sptOFTN}: These sources should present nearly complete Einstein rings at higher resolution. 
In the case of \sptOFOE, this is due to the excellent alignment of source and tangential caustic. \sptOFTN\ is the most extended source 
in this sample and its large size compared to the caustic fills the ring effectively. In both cases, the luminosities, submillimeter fluxes, and source sizes 
are very comparable to those observed in other samples of distant starburst galaxies \citep[e.g.,][]{rujopakarn11}. The posterior of the model parameters for \sptOFTN\ show two separate peaks in the lens and source positions, but the magnifications of both models are similar.  The extended array data on this source can possibly break this degeneracy.

\begin{figure*}[ht]
\includegraphics[trim = 30mm 0mm 0mm 0mm, width=1.00\textwidth]{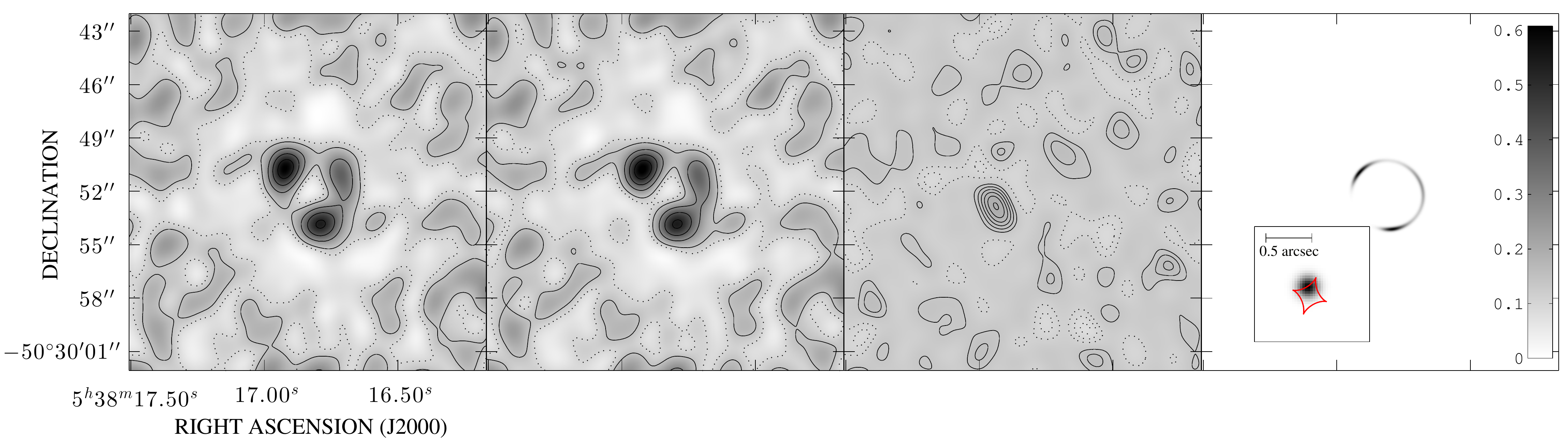}
\caption{\label{fig:0538OneBlob} The best fit lens model for SPT0538-50 using a single source component, for comparison with 
Figure~\ref{fig:modeling}. Left to right are the ``dirty'' image, the model dirty image, 
the difference between data and model, and the full-resolution source 
model. Contours in the two panels on the left start at $\pm$2 times the rms noise in the residual map ($\sigma$), 
increasing in steps of 10$\sigma$, while contours in the third panel increase  
in steps of $\pm$1$\sigma$. The peak in the residual map has a significance of $>$5$\sigma$. As shown in Figure~\ref{fig:modeling}, 
a second source component completely eliminates this high peak and has a 
significance $>$10 $\sigma$ in the visibility space. 
}
\vspace{1mm}
\end{figure*}

{\it \sptOFTE}: The large angular size of this ring suggests a massive lens, with the models indicating that the lens has a projected mass of
nearly $10^{12}$~\msol\ within a radius of 10~kpc. 
Close alignment between the compact source (labeled component ``A''  in Table~\ref{table:sources}), and caustic again leads to a large magnification. A single-component source does not 
provide an adequate fit,  leaving a significant residual structure to the south east of the lens center (Figure~\ref{fig:0538OneBlob}). 
A far better match to the data is the model shown in Figure~\ref{fig:modeling} (bottom), which includes a second source component (labeled component ``B'' in Table~\ref{table:sources}), 
 offset from and much more extended than the first, and representing 
30\% of the total source-plane luminosity.

\section{Conclusion}
We have used ALMA to image the submillimeter emission from four DSFGs discovered by the SPT. We find that these objects
resolve into ring-like structures expected from gravitational lensing, a picture confirmed by the redshift information we have for 
the submillimeter emission and NIR counterparts. We present a visibility modeling procedure to fit gravitational lens models to 
these data and simultaneously correct the unknown phase errors of the antennas introduced by, e.g., imperfect antenna positions.
From this technique, we are able to correct for the magnification of the sources presented in this work and derive intrinsic properties, finding the galaxies
to be typical high-redshift, DSFGs. The sensitivity of ALMA permits these lens models to be constrained 
in short observations. Longer observations of lensed starbursts in future cycles will therefore enable studies of ISM structure and lower 
luminosity molecular lines that are otherwise impossible to observe in unlensed systems.

\acknowledgements{The SPT is supported by the National Science Foundation
through grant ANT-0638937, 
with partial support through PHY-1125897, the Kavli Foundation and the Gordon and Betty 
Moore Foundation. 
This paper makes use of the following ALMA data: ADS/JAO.ALMA \#2011.0.00957.S and \#2011.0.00958.S. ALMA is a partnership of ESO (representing its member states), NSF (USA) and NINS (Japan), together with NRC (Canada) and NSC and ASIAA (Taiwan), in cooperation with the Republic of Chile. The Joint ALMA Observatory is operated by ESO, AUI/NRAO and NAOJ.
The National Radio Astronomy Observatory is a facility of the National Science Foundation operated under cooperative agreement by Associated Universities, Inc. 
Partial support for this work was provided by NASA through grant HST-GO-12659 from the Space Telescope Science Institute 
and an award for {\it Herschel} analysis issued by JPL/Caltech for OT2\_jvieira\_5. 
Work at McGill is supported by NSERC, the CRC program, and CIfAR. YDH acknowledges the support of FQRNT through International Training Program and Doctoral Research scholarships. 
TRG acknowledges support from the Science and Technologies Facilities Council.}

\appendix
\section{Lens Model Self-Calibration}
Interferometric phase calibration procedures generally leave some residual phase errors due to imperfect baseline solutions, 
uncompensated atmospheric delays, or other effects. The magnitude of these errors depends on many factors, such as calibration interval 
and calibrator-source separation, and their importance depends on the signal to noise ratio in the data and the complexity of the imaging task. 
In the present application, small phase errors can redistribute flux between lensed images and without properly accounting for such effects in 
our lens modeling, the derived model parameter distributions may be significantly in error. 

A standard procedure to correct antenna-based phase errors is self-calibration. Using a source model derived from images of the corrupted 
visibilities, phase corrections are derived, the source is re-imaged, and another iteration can be made using a new source model produced from the phase-corrected visibilities. 
For the present purposes, the most significant disadvantage of this phase correction scheme is that the uncertainties associated with the 
source model against which the data are self-calibrated are not included in the lens modeling, and any structural errors introduced by noise 
or phase errors may become a permanent part of the lens model. 
Here we propose a method to incorporate the phase correction into the lens modeling procedure, using the lensed structure as the source 
model for self calibration. 

To implement this self calibration technique, we use a perturbative approach in which we assume that the current data is equal to the model 
plus an antenna based phase corruption. We can write the expression for $\chi^2$ for the visibility phases between the data and model as
\begin{equation}
\chi^2 = \left[\delta \Phi_i +\delta \phi_k \frac{\partial \Phi_i}{\partial \phi_k}\right] C^{-1} \left[\delta \Phi_i + \frac{\partial \Phi_i}{\partial \phi_k} \delta \phi_k \right]
\end{equation}
where $\Phi_i$ is the phase of the $i$'th visibility, $\delta \Phi_i$ is the phase difference between the model and the data and 
$\delta \phi_k$ is the phase delay in the $k$'th antenna. 
$\partial \Phi_i /\partial \phi_k$ is a Jacobian matrix containing the gradients of the observed visibilities with respect to changes in antenna phases. For N antennas and M visibilities $\partial \Phi /\partial \phi$ is an $N \times M$ matrix whose $ik$'th element is 1 if the first 
antenna of $i$'th visibility is $k$, $-1$ if the second antenna of $i$'th visibility is $k$ and zero otherwise.
C is a diagonal covariance matrix whose nonzero elements are 
approximately $(\sigma_{xy} / |V|)^2$, where $\sigma_{xy}$ is the rms error on the real or imaginary part of each visibility and $|V|$ is 
the visibility amplitude. Note that this approximation to the covariance matrix is only valid in the limit of high signal to 
noise data \citep{wrobel99}. 

\begin{figure}[h]
\includegraphics[width=1.00\textwidth]{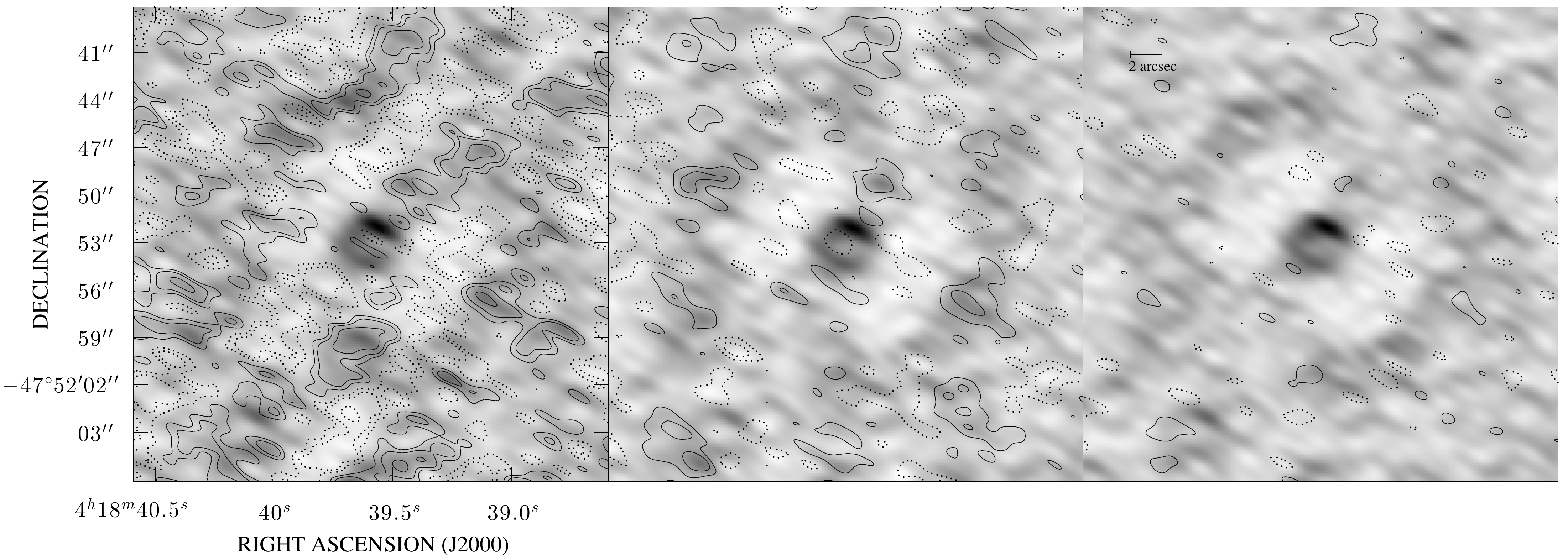}
\caption{\label{fig:selfcal}
A comparison of image residuals for three different corrections to the antenna phases. The dirty image of \sptOFOE\ is shown in greyscale beneath each panel. Contours show the image residuals after the source model is subtracted from the dirty image. 
\textbf{Left:} Residual structure after the model that fits the raw data (no self-calibration) best is subtracted from the data. \textbf{Middle:} The same, 
but for data calibrated using a CLEAN-based self-calibration. \textbf{Right:} The same, but for data calibrated using the procedure described in this work.
The presence of residual structure away from the source, with no corresponding residual at the source position, is a clear sign 
of imperfectly corrected phase in the left two panels.}
\end{figure}

To minimize $\chi^2$ we set its derivative with respect to antenna phases to zero ($\partial \chi^2 / \partial \phi_i=0$). This allows us to write the antenna phase offsets as
\begin{equation}
\delta \phi_l = -(F^{-1})_{lk} \frac{\partial \Phi}{\partial \phi} \, \, (C^{-1})_{ji} \, \,  \delta \Phi_i~,
\label{eq:dphi}
\end{equation}
 where F is the Fisher matrix calculated as
\begin{equation}
F_{ij}=\frac{\partial \Phi_k}{\partial \phi_i} (C^{-1} )_{kl} \frac{\partial \Phi_l}{\partial \phi_j}~.
\end{equation}
 At every iteration of the MCMC code, the value of $\chi^2$ is minimized for the postulated model by deriving calibration phases using equation~\ref{eq:dphi}. The resulting $\chi^2$ is used to evaluate the likelihood and determine the next link in the chain, thereby 
incorporating the uncertain phase correction in the parameter exploration. In the limit of an intrinsically Gaussian distribution for these phase parameters, this is equivalent to marginalizing over the phases.  In Figure~\ref{fig:selfcal} we compare the results of this procedure with 
the standard self-calibration based on CLEAN components, and the raw data. The improvement in source subtraction is significant, even compared to the standard CLEAN procedure.

\begin{figure}[h]
\includegraphics[width=1.00\textwidth]{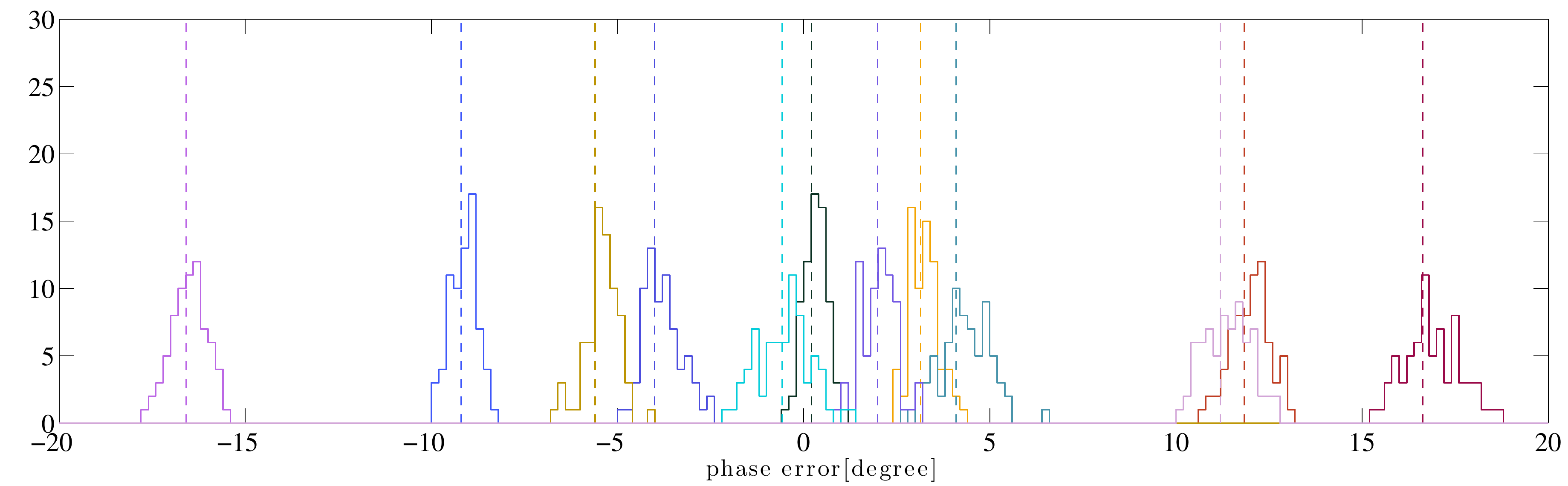}
\caption{\label{fig:phase}
Antenna phase errors recovered in 70 simulated observations of a lens model with different noise realizations. Phase errors were introduced to the antennas, with values indicated by the vertical dashed lines. The histograms show the phase values recovered in fits to the simulated visibilities following the technique outlined here. }
\end{figure}

 A second test of the simultaneous fitting of the lensed emission and antenna phases is shown in Figure~\ref{fig:phase}. 
Simulated observations of a typical 
lens model were created with realistic noise levels, and antenna phase errors added to these visibilities. The MCMC fitting algorithm was applied 
to these multiple realizations of the same source to verify that the antenna phase errors are recovered. The excellent agreement 
between input and recovered phase demonstrates that the simultaneous fitting of the lens model and antenna phases 
does not bias the antenna phase measurement, despite the complicated source structure of the lens models. 

\newpage 

\newpage 

\section{High Resolution Observations of SPT0346-52}
\begin{figure}[h]
\center{\includegraphics[width=0.70\textwidth]{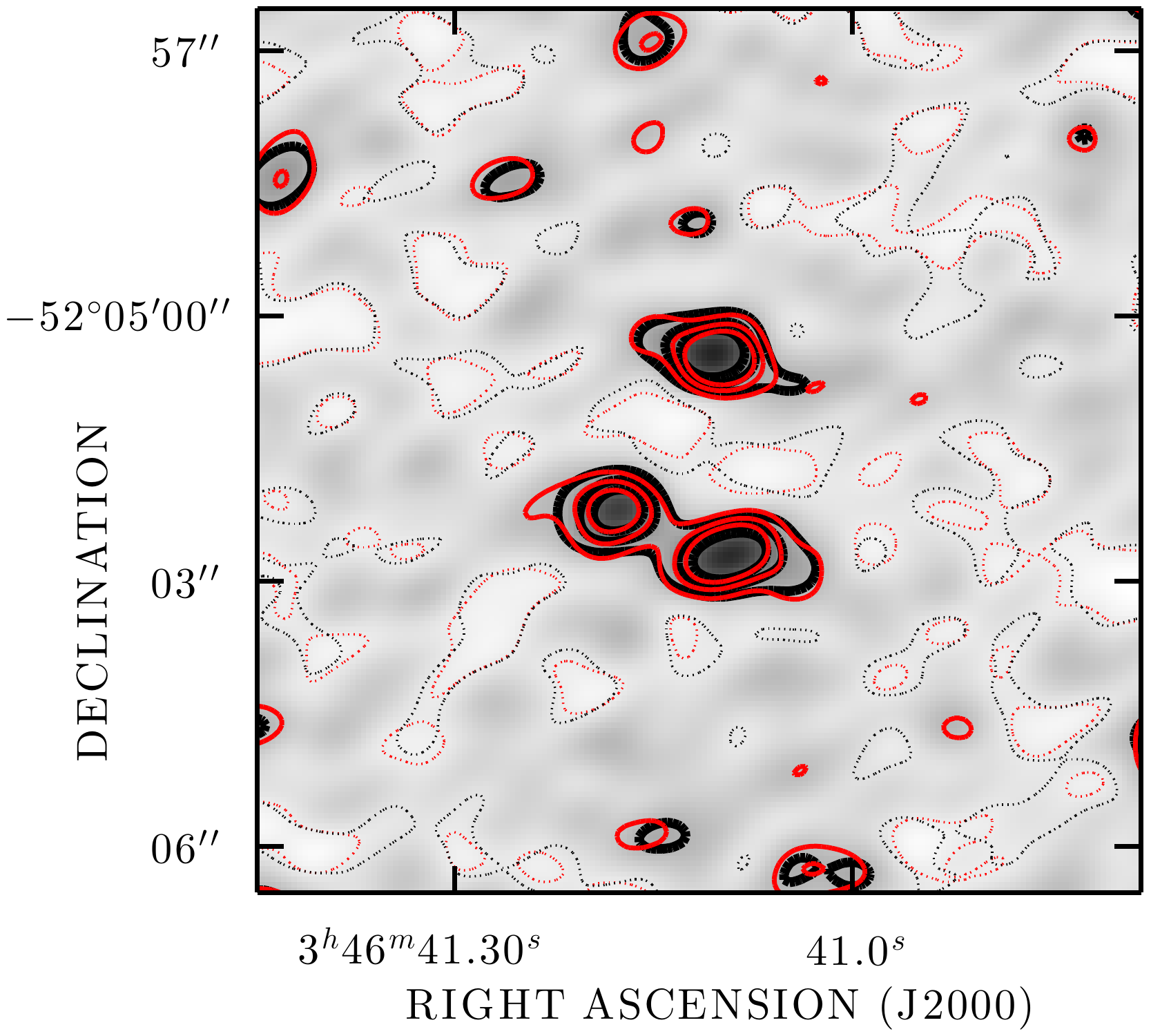}}
\caption{\label{fig:extended} A comparison of the model proposed for \sptOTFS\ from the low-resolution observations presented above 
and the higher-resolution observations obtained after submission. The ``dirty'' image of the extended configuration observations 
are shown in greyscale and black contours. The red contours are the predicted appearance given the model of Figure~\ref{fig:modeling} and the 
$uv$ sampling of the new data. Contours are drawn at -5 (dashed) and 20, 40, 60 (solid) times the rms noise.}
\end{figure}

The models in Figure \ref{fig:modeling} were derived using the compact configuration data that were available at the time of submission. Higher resolution data were delivered later, and permit a direct comparison of the model based on the low-resolution data with the higher resolution observations of \sptOTFS. Figure \ref{fig:extended} shows the predictions of the best fit model (from fitting to the compact data, presented earlier in this work) for the uv-coverage of the extended data (red contours). The black contours show the extended data. The contours demonstrate a high degree of agreement between the predictions and the new observations.
A lens model for the extended configuration data shows a consistent model, resulting in magnification (from modeling the extended data alone) of  $5.26 \pm 0.12$ in agreement with the magnification derived from the compact data.

\bibliographystyle{apj}

\end{document}